\newcommand{\diracslash}[1]{#1\llap{/\kern2pt}}

\newcommand{\be}{\begin{equation}}
\newcommand{\ee}{\end{equation}}
\newcommand{\bea}{\begin{eqnarray}}
\newcommand{\eea}{\end{eqnarray}}
\newcommand{\ba}[1]{\begin{array}{#1}}
\newcommand{\ea}{\end{array}}

\newcommand{\bt}{\begin{tabular}}
\newcommand{\et}{\end{tabular}}

\newcommand{\beas}{\begin{eqnarray*}}
\newcommand{\eeas}{\end{eqnarray*}}

\documentclass[amsmath,11pt,amssymb,preprint,nofootinbib]{revtex4}
\usepackage{amsfonts} 
\usepackage{graphicx} 
\usepackage{epsfig}
\usepackage{bm}

\begin{document}

\title{Strange mesons in strong magnetic fields} 

\author{Amruta Mishra}
\email{amruta@physics.iitd.ac.in}
\affiliation{Department of Physics,Indian Institute of Technology,Delhi,
Hauz Khas,New Delhi--110 016,India}

\author{S.P. Misra}
\email{misrasibaprasad@gmail.com}
\affiliation{Institute of Physics, Bhubaneswar -- 751005, India} 

\begin{abstract}
The masses of the strange mesons ($K$, $K^*$ and $\phi$) 
are investigated in the presence of strong magnetic fields. 
The changes in the masses of these mesons arise from the mixing 
of the pseusdoscalar and vector mesons in the presence 
of a magnetic field. For the charged mesons, these mass modifications
are in addition to the contributions from the lowest Landau 
energy levels to their masses. 
The decay widths, $\phi \rightarrow K\bar K$ and $K^* \rightarrow K\pi$,
in the presence of the magnetic field are studied using a field theoretic
model of composite hadrons with constituent quarks/antiquarks. The model
uses the free Dirac Hamiltonian in terms of the constituent quark 
fields as the light quark antiquark pair creation term and explicit 
constructions for the meson states in terms of the constituent
quarks and antiquarks to study the decay processes. 
The pseudoscalar-vector meson (PV) mixing leads to 
a drop (rise) in the mass of the pseudoscalar (longitudinal component of 
the vector meson). 
It is observed that the mass modifcations 
for the hidden strange mesons, arising from $\phi-\eta'$ mixing, 
are quite prominent, whereas the neutral open strange mesons 
have only marginal changes in their masses due to mixing. 
The mass modifications of the charged open strange mesons, 
are due to interplay between the Landau level contributions 
and the PV mixing, and, the latter effect is observed to dominate 
at large values of the magnetic field.
The vector meson decay widths ($\phi\rightarrow K\bar K$ and
${K^*}\rightarrow K\pi$) involving the charged mesons are 
observed to be to be quite different from the widths involving 
neutral mesons, due to the additonal contribution 
for the charged mesons from the Landau levels. 
\end{abstract}

\maketitle

\def\bfm#1{\mbox{\boldmath $#1$}}
\def\bfs#1{\mbox{\bf #1}}

\section{Introduction}
The topic of the in-medium properties of strange hadrons is 
a subject of intense research in the context of the heavy 
ion collision experiments as well as in the study
of the bulk matter in nuclear astrophysical objects, e.g., 
(proto) neutron stars \cite{Tolos_Prog_Part_Nucl_Phys_112_103770_2020}. 
The topic is important due to its
relevance to the experimental observables, e.g., the collective 
flow as well as the yield and spectra of these hadrons 
resulting from the high energy nuclear collision experiments
\cite{C_Hartnack_Phys_Rep_510_2012_119}. 
The strange baryons could exist in the high density bulk
matter in neutron stars and there could be a 
possibility of the antikaon condensation \cite{kaplan} 
in the interior of the neutron stars.

There have been extensive studies of the $K$ and
$\bar K$ mesons in the literature.
In the early work on the study of kaons and antikaons
based on chiral perturbation theory \cite{kaplan},
the masses of these mesons were obtained by solving 
the dispersion relation obtained by using the leading 
order Weinberg-Tomozawa term 
($\sim(\bar N \gamma^\mu N) (\bar K (\partial_\mu K) - 
(\partial_\mu \bar K) K$)) and the next to leading order
$KN$-sigma term ($\sim\Sigma_{KN} (\bar N N)(\bar K K)$).
The leading term gives a rise (drop) in the mass
of the kaon (antikaon). On the other hand, 
the second term is attractive for both kaons 
and antikaons and is largely responsible for the 
possibility of the $K^-$ condensation 
in the interior of a neutron star \cite{kaplan}.
Using a meson exchange model
\cite{Glendenning_Schaffner_PRC_60_025803_1999,Debades_PRC86_045803_2012},
the $K$ and $\bar K$ mesons in hadronic matter 
have been studied, where their interactions to the meson fields 
arise in the same manner as the baryon-baryon interactions
(via exchange of $\sigma$, $\omega$, $\rho$ mesons) 
in the relativistic Quantum Hadrodynamics (QHD) model
\cite{Serot_QHD}. 
The interaction due to scalar meson exchange 
is attractive for both kaons and antikaons leading 
to a mass drop of these mesons in the medium,
similar to the mass shift of the baryons
in QHD, which arises due to the scalar potential
experienced by the baryon in the hadronic medium. 
The open strange mesons have also been studied
using the Quark meson coupling (QMC) model.
In the QMC model, the light ($u$, $d$) quark (antiquark) 
constituents of the hadron, interact via the exchange 
of the light mesons ($\sigma$, $\omega$ and $\rho$) 
and the mass shifts of the hadrons
arise dominantly due to the scalar potentials experienced by 
these constituents. The hadrons,
with the same light quark and antiquark constituents,
e.g., $K$ and $K^*$ mesons,
are thus observed to have very similar mass shifts
\cite{Krein_Prog_Part_Nucl_Phys,qmc_phi}.
The vector potential felt by a hadron, $h$, in the medium
is given as $U_v^h=(n_q-n_{\bar q})V_\omega^q+I_3^h V_\rho^q)$,
where $n_q$ and $n_{\bar q}$ refer to the number of light quarks
and antiquarks present inside the hadron, and,
$I_3^h$ is the third component of the isospin 
of the hadron. It might be noted here that, the vector potential 
for the non-strange quark (antiquark), $\pm V_\omega^q$ 
in $K(\bar K)$ meson in the Quark meson coupling (QMC) model
has to be increased by a factor $(1.4)^2$ 
to reproduce the repulsive $K^+$--Nucleus interaction
\cite{qmc_phi}.

The coupled channel approach has been used
extensively in the literature to study 
the spectral functions of the $K$ and $\bar K$ mesons. 
Within this approach, the $\Lambda (1405)$ 
resonance is generated dynamically from the $\bar K N$ interaction,
using the lowest order meson-baryon interaction in a chiral 
Lagrangian and a cut-off to regularize the loop integrals
\cite{Oset_RamosNPA635_1998_99}. The spectral function
of $\bar K$ in the medium is obtained from the
$\bar K N$ scattering amplitude, $T_{\bar KN}$
by a self consistent solution of the coupled channel
Lippmann Schwinger equations. The modifications
of the masses in hot hadronic matter include the effects
from the Pauli blocking, the mean field binding potentials 
for the baryons, as well as, the self-energies of the $\bar K$ 
and $\pi$ \cite{Oset_Ramos_NPA_671_481_2000}.
For the $K N$ interaction, due to absence of any resonance
near to the threshold, the $T\rho$ approximation is a good approximation 
for the study the self-energies of the kaons 
at low densities, and, a self-consistent calculation of 
the scattering amplitude has only small modifications 
to results obtained using this approximation.
The medium modifications of the spectral functions
of the $K$, $\bar K$, as well as, of the vector strange 
mesons, $K^*$ and $\bar {K^*}$, 
calculated using the self-consistent coupled channel approach, 
have been used to study the dynamics of $K^*$ and $\bar {K^*}$
in heavy ion collision experiments, 
using the Parton-Hadron-String-Dynamics (PHSD) transport model,
which incorporates partonic as well as hadronic degrees
of freedom \cite{Elena_16,Elena_17}. In Ref. \cite{Elena_16},
the PHSD calculations for Au+Au collisions 
at $\sqrt {s_{NN}}$=200 GeV are observed to be in good agreement 
with the data from STAR Collaboration at RHIC \cite{STAR_Kstr}. 
These PHSD calculations show that the production of $K^*$($\bar {K^*}$) 
by hadronization from the QGP phase 
are quite small and these vector mesons are dominantly
produced at the later hadronic stage from $K(\bar K)\pi$ 
scatterings. The medium effects on the spectral 
functions of these mesons are observed to be quite small
due to the small densities and temperatures attained 
by the medium when the $K^*$ and ${\bar {K^*}}$ mesons 
are produced. The PHSD calculations for Pb+Pb collisions
at $\sqrt {s_{NN}}$=2.76 TeV for the particle spectra
and particle ratios \cite{Elena_17} have been 
compared to the experimental data obtained 
by the ALICE Collaboration at the LHC
\cite{ALICE_Kstr}. These calculations show that
at LHC too, the main source of the $K^*$ and ${\bar {K^*}}$
are from the later hadronic stage from $K(\bar K)\pi$ scattering,
and the production from from the QGP phase
by hadronization is very small. In \cite{Elena_17}, the PHSD calculations
have also been performed for heavy ion collision experiments 
at low beam energies relevant for the FAIR and NICA conditions, 
where the medium effects may become visible due to the
large baryon density matter which will be created at these
facilities. 

The kaons and antikaons in hadronic matter have been studied 
using a chiral effective model based on a non-linear realization
of chiral symmetry and broken scale invariance \cite{paper3}.
The in-medium masses of $K$ and $\bar K$ have been studied 
retaining the leading order Weinberg Tomozawa term,
as well as, the next to leading 
order terms (the scalar exchange and the range terms)
in the chiral perturbative expansion, consistent
with the kaon-nucleon scattering lengths 
\cite{kmeson1,isoamss,isoamss1,isoamss2}. 
The model has been used recently to study the 
in-medium decay widths of the $\phi$, $K^*$ and $\bar {K^*}$
to their dominant decay modes of $K\bar K$, $K\pi$
and $\bar K \pi$ respectively, from the mass modfications
of the open strange mesons, and,
the effects of the isospin asymmetry and the strangeness
fraction of the strange hadronic medium on the spectral functions 
and the production cross-sections of the vector mesons, 
$\phi$, $K^*$, ${\bar {K^*}}$ have also been studied 
\cite{strangedecaywidths}.

Recently, there have been a lot of studies 
on the properties of the hadrons in the presence
of strong magnetic fields.
This is due to the estimation of the magnetic fields produced 
in the peripheral ultra relativistic heavy ion collision experiments 
to be huge \cite{Tuchin_Review_Adv_HEP_2013}.
Also, strong magnetic fields exist in the astrophysical 
compact objects, e.g. magnetars, where the  magnetic field
is of the order of $10^{15}-10^{16}$ Gauss at the surface 
and the magnitude of the magnetic field could be much larger
in the interior of magnetars \cite{magnetars}.
The created magnetic field has been estimated
to be of the order of $eB\sim 5 m_\pi^2$ 
for Au-Au collisions at $\sqrt {s_{NN}}$=200 GeV at RHIC 
\cite{Kharzeev_NPA803_227_2008,Skokov_Illarionov_Tonnev_IJMPA24_5925_2009}
and $eB \sim  15 m_\pi^2$ at LHC
\cite{Skokov_Illarionov_Tonnev_IJMPA24_5925_2009}.
It was suggested that in the presence of strong magnetic 
fields in heavy ion collision experiments, the chiral 
magnetic effect (CME) can lead to preferential emission 
of charged particles along the direction of the magnetic field
\cite{Kharzeev_NPA803_227_2008,Kharzeev_PRD78_074033_2008}.
An electric charge asymmetry of produced particles above 
and below the reaction plane has been observed by the STAR 
Collaboration at RHIC \cite{STAR_CME}, which could be 
due to the separation of charge along the direction of the
magnetic field resulting from the chiral magnetic effect. 
The creation and evolution of a magnetic field
in heavy ion collision experiment employed in Hadron String 
Dynamics (HSD) for non-central Au-Au collisions 
at $\sqrt {s_{NN}}$=200 GeV \cite{Voronyuk}, however,
does not show much effect on the experimental observables 
due to the presence of the electromagnetic field.

As has already been mentioned, the magnetic fields
resulting from the non-central ultra-relativistic 
heavy ion collisions at Relativistic Heavy Ion Collider (RHIC) at BNL 
and Large Hadron Collider (LHC) at CERN are estimated
to be huge. The magnetic fields created from the heavy 
ion collisions, drop rapidly after the collision. 
In the presence of a medium, the drop
in the magnetic field leads to induced currents 
($\sim \sigma {\bf E}$, with $\sigma$ as the electrical conductivity
of the medium) being created, which tend to slow down the decay 
of the magnetic field. However, the time evolution of the magnetic field
\cite{Tuchin_Review_Adv_HEP_2013} 
is still an open problem and it needs 
the solution of the magnetohydrodynamic equations \cite{Ajit_MHD},
with a proper estimate of the electrical conductivity 
of the medium.

The effects of a magnetic field on the light mesons,
e.g., $\pi$, $\rho$, $K$ and $\phi$ mesons,
have been studied in the literature
\cite{Aguirre_light_mesons,Aguirre_neutral_K_phi,Pradip_rho_BT,Chernodub}.
In the presence of an ultra-strong magnetic field, $B>B_{crit}$,
where, $B_{crit}={m_\rho}^2/e$, where $m_\rho$
is the mass of the $\rho$ meson, the interesting possibility
of vacuum superconductivity with condensation of the charged 
$\rho$ mesons has been suggested \cite{Chernodub}. 
Alsom the effects of magnetic field on the strange mesons
$\phi$ and neutral kaons have been studied 
in Ref. \cite{Aguirre_neutral_K_phi}
for the neutron star matter for a range of magnetic field
($10^{15}-10^{19}$ Gauss). The study shows
an oscillatory behaviour around the vacuum value
for the decay width of $\phi \rightarrow K^+ K^-$
for small values of the magnetic field,
and a critical field ($B_{crit}\sim 2 \times 10^{18}$ Gauss)
above which the decay width vanishes. This is 
due to the positive Landau level contributions 
to the masses of the charged kaons and antikaons 
in the presence of a magnetic field.
In the case of a large magnetic field 
in the interior of the neutron star, 
the increase in the mass of $K^-$ due to Landau contributions
could have implications on the possibility of $K^-$ condensation
inside the neutron star.

There have been a lot of studies on the heavy flavour mesons 
in the recent past \cite{Hosaka_Prog_Part_Nucl_Phys} and
in the presence of a magnetic field, 
the mixing of the pseudoscalar and the vector mesons
\cite{charmonium_mag_QSR,charmonium_mag_lee,Gubler_D_mag_QSR,Alford_Strickland_2013,Suzuki_Lee_2017} 
has been shown to lead to dominant modifications
to the masses of these mesons.
The study of heavy flavour charmonium states as well as 
of $D$ mesons in the presence of a magnetic field
within a QCD sum rule approach 
\cite{charmonium_mag_QSR,charmonium_mag_lee,Gubler_D_mag_QSR}
as well as within an effective potential approach
\cite{Alford_Strickland_2013} show the effects of the
mixing on the masses to be quite pronounced.
A study of the mixing effects on the formation times
of the vector and pseudoscalar charmonium states
($J/\psi-\eta_c$ and $\psi'-\eta'_c$ mixing effects) 
shows to lead to faster (slower) formation times for the
pseudoscalar (vector) states \cite{Suzuki_Lee_2017}.
Due to the mixing with the vector charmonium states, 
the pseudoscalar mesons, $\eta_c$ and $\eta'_c$ might show as 
peaks in the dilepton spectra,
arising from anomalous decay modes $\eta_c,\eta'_c \rightarrow
l^+l^-$ amd can act as probe of a strong magnetic field
at the early stage \cite{Suzuki_Lee_2017}. 

The changes in the masses of the charmonium states 
\cite{charmonium_PV_amspm}
and the open charm mesons \cite{dmeson_PV_amspm}
in the presence of strong magnetic fields
due to the pseusoscalar-vector mixing effects,
in addition to the Landau level contributions
to the charged mesons, are observed to be quite appreciable. 
The effects of the magnetic field on the decay widths 
$\psi(3770)\rightarrow D\bar D$ \cite{charmonium_PV_amspm}
as well as $D^* \rightarrow D\pi$ \cite{dmeson_PV_amspm}
are calculated from their
mass modifications in the presence of a magnetic field,
using a field theoretical model of composite hadrons with 
constituent quark (antiquark) \cite{spm781,spm782}. 
The model of composite hadrons uses the light quark-antiquark
pair creation term of the free Dirac Hamiltonian in terms
of the constituent quark fields and explicit constructions
of the initial and final states to study the decay processes. 
In the absence of a magnetic field, the model has been used to
calculate the decay widths of the charmonium states to $D\bar D$
and $D^*\rightarrow D\pi$ \cite{amspmwg}, as well as, 
to study the decay widths of bottomonium states to $B\bar B$
\cite{amspm_upsilon} in asymmetric strange hadronic matter.
The charmonium decay widths to $D\bar D$ were also calculated
using a light quark antiquark pair creation
(in the $^3P_0$ state) model \cite{friman,amarvepja}, 
namely, the $^3P_0$ model \cite{3p0,3p0_1}.
The modifications of the decay widths were computed 
from the mass modifications of the charmonium states
as well as $D$ and $\bar D$ mesons 
calculated using a chiral effective model \cite{amarvepja}. 
When the internal structure of the mesons in the initial and
final states are taken into account, within the composite model
of hadrons \cite{amspmwg}, as well as using the $^3P_0$ model \cite{friman},
the heavy quarkonium decay widths were observed to vanish
at specific densities (so called nodes) \cite{friman,amarvepja}.
In both the models, the decay process is through the 
creation of a light quark-antiquark pair 
and the constituent quark (antiquark) of the decaying meson,
combines with the light antiquark (quark) created to form the
final state mesons.
The study of the heavy quarkonium decay widths
in asymmetric hyperonic matter 
using the field theoretical model of composite hadrons 
\cite{amspmwg,amspm_upsilon}
show that the effects of density are the dominant medium effects
as compared to the effects from isospin asymmetry
as well as strangeness.
The mass modifications of the hidden and open charm mesons
in the presence of strong magnetic fields,
arising from the pseudoscalar-vector mesons mixings,
in addition to the Landau contributions to the masses
of the charged mesons, and their effects on the decay 
widths of charmonium states
to $D\bar D$ and $D^*\rightarrow D\pi$ 
\cite{charmonium_PV_amspm,dmeson_PV_amspm} 
have been studied using the  model of composite hadrons
\cite{spm781,spm782,spmdiffscat}. 
In the present work, we study the mass modifications 
of the strange $K$, $K^*$, $\phi$ mesons
in the presence of a magnetic field
due to the mixing of the pseudoscalar ($P$) and vector ($V$) 
mesons ($K-K^*$ and $\phi-\eta'$ mixing),
including the Landau level contributions
for the charged mesons, and their subsequent 
effects on the decay widths $K^* \rightarrow K \pi$ and 
$\phi \rightarrow K \bar K$. 

The outline of the paper is as follows. In section II,
the mass modifications of the strange $K$, $K^*$ and $\phi$ mesons 
and their effects on the decay widths $K^* \rightarrow K\pi$
and $\phi \rightarrow K \bar K$,
are investigated in the presence of a uniform magnetic field.
The mass modifications arise from the pseudoscalar-vector mesons
mixing ($K-K^*$ and $\phi-\eta '$) 
in the presence of a magnetic field. 
For the charged $K$ and $K^*$ mesons, the Landau level contributions 
to these masses are also taken into account.
The effects of the mass modifications 
on the decay widths of $K^*\rightarrow K\pi$ as well as
of $\phi$ meson to $K \bar K$ ($K^+K^-$ and $K^0 \bar {K^0}$)
are investigated from the mass modifications of these mesons
in the presence of a magnetic field.
These decay widths are studied using a field theoretical
model of composite hadrons with quark and antiquark 
constituents. 
We discuss the results of the masses and
decay widths of the strange mesons in the presence of 
strong magnetic fields in section III. 
Section IV summarizes the results of the present study.

\section{Strange mesons in presence of a magnetic field}
\subsection{MASSES:}
In this section, the mass modifications for the $K$ and
$K^*$ mesons, as well as of $\phi$ and $\eta '$,
are studied arising due to the pseudoscalar-vector meson mixing
in the presence of a magnetic field.
The effect of the mixing is considered through a phenomenological
Lagrangian interaction \cite{charmonium_mag_lee,Gubler_D_mag_QSR}
\begin{equation}
{\cal L}_{PV\gamma}=\frac{g_{PV}}{m_{\rm {av}}} e {\tilde F}_{\mu \nu}
(\partial ^\mu P) V^\nu,
\label{PVgamma}
\end{equation}
where $m_{\rm {av}}$ is the average of the masses
of the pseudoscalar and vector mesons.
The coupling strengths of the radiative decay of the 
vector meson, $V$ to the pseudocalar meson, $P$, as
described by the interaction given by (\ref{PVgamma}) are
determined from the observed decay widths of $V\rightarrow P\gamma$ 
in vacuum. 

For the charged $K$ and $K^*$ mesons,
the contributions of the mixing on the masses of these mesons
are in addition to the lowest level Landau level contributions
to their masses. The masses of the charged pseudoscalar 
and vector mesons in their ground states in a background 
magnetic field are given as \cite{Chernodub}
$m_{P} (B)=(m_{P}^2+ eB)^{1/2}$ and 
$m_{V} (B)=(m_{V}^2 - eB)^{1/2}$,
with $m_{P,V}$ as the masses
of these mesons at zero magnetic field. 
For the vector meson, the gyromagnetic ratio is taken to be
2 \cite{Gubler_D_mag_QSR,Chernodub}.
For the neutral mesons, $m_{V,P}(B)=m_{V,P}$, 
as there are no Landau level contributions.
In the presence of a magnetic field, there is
mixing of the pseudoscalar and the longitudinal component 
of the vector mesons, which modify their masses to 
\begin{equation}
m^{(PV)}_{V^{||},P}=\Bigg [\frac{1}{2} \Bigg ( M_+^2 
+\frac{c_{PV}^2}{m_{\rm {av}}^2} \pm 
\sqrt {M_-^4+\frac{2c_{PV}^2 M_+^2}{m_{\rm {av}}^2} 
+\frac{c_{PV}^4}{m_{\rm {av}}^4}} \Bigg) \Bigg]^{1/2},
\label{mpv_long}
\end{equation}
where $M_{\pm}^2=m^2_V(B) \pm m^2_P (B)$, 
$c_{PV}= g_{PV} eB$ and $m_{\rm {av}}=(m_V (B) +m_P(B))/2$. 

\subsection{DECAY WIDTHS:}
The decay widths of $K^* \rightarrow K \pi$ and $\phi \rightarrow 
K\bar K$ are modified in the presence of a magnetic field 
due to the changes in the masses of these mesons arising 
from the pseudoscalar-vector mesons mixing effects, in addition
to the Landau level contributions for the charged mesons.
These decay widths are studied using a field theoretical
model of composite hadrons with quark/antiquark constituents
\cite{spm781,spm782,spmdiffscat}.
The matrix element of the quark antiquark pair creation term
of the free Dirac Hamiltonian term for light quarks ($q=u,d)$
is evaluated between the initial and final
meson states to calculate the decay widths.
The initial and final meson states are constructed
explicitly in terms of the constituent quark fields, assuming 
the harmonic oscillator potential between the quark and
antiquark constituents.
Using a Lorentz boosting, the constituent quark field operators 
of the hadron in motion are obtained from the constituent quark 
field operators of the hadron at rest.

We investigate the decay processes, 
${K^*}^+ \rightarrow K\pi (K^+\pi^0,K^0 \pi^+)$,
${K^*}^0 \rightarrow K\pi (K^0\pi^0, K^+\pi^-)$
and $\phi \rightarrow K\bar K (K^+K^-,K^0 \bar {K^0})$,
within the composite model of hadrons in the present work.
For the vector mesons, $K^*$ and $\phi$ decaying 
at rest, the outgoing pseudoscalar meson 
($K$, $\pi$, $K$, $\bar K$) is with finite momentum.
The explicit construction for the final state pseudoscalar 
meson, $P$ with momentum ${\bf p}$,
assuming harmonic oscllator wave function,
is given as
\begin{equation}
|P ({\bf p})\rangle  = 
\frac{1}{\sqrt{6}}
\Bigg (\frac {R_P^2}{\pi} \Bigg)^{\frac{3}{4}}
\int d{\bf k} 
\exp \Big(-\frac {R_P^2 {\bf k}^2}{2}\Big)
{Q_1}_r^i({\bf k}+\lambda_2 {\bf p})
^\dagger u_r^\dagger 
{\tilde {Q_2}}_s^i 
(-{\bf k} +\lambda_1 {\bf p})v_s
|vac\rangle,
\label{pseudoscalar_meson}
\end{equation}
where, ${{Q_1}_r^i}({\bfs k})^\dagger ( {{\tilde {Q_2}}_r^i}({\bfs k}))$
is the quark (antiquark) creation operator of flavour $Q_1 (Q_2)$
with spin $r$, color $i$ and momentum ${\bf k}$,
and $u_r$ and $v_s$ are the two component spinors.
For the meson $P(\equiv K^+,K^0, K^-,\bar {K^0}, \pi^+,\pi^-,\pi^0)$, 
the quark-antiquark constituents are given as   
$(Q_1,{\bar {Q_2}})\equiv (u,\bar s), (d,\bar s), (s,\bar u), (s,\bar d), 
(u,\bar d), (d,\bar u),
 \frac {1}{\sqrt 2}\Big[(u,\bar u) -(d, \bar d)\Big]$.

The decaying vector meson, $V(\equiv {K^*}^+, {K^*}^0, \phi$) 
at rest, with polarisation $m$, is constructed as \cite{amspmwg}
\begin{eqnarray}
|{V}^m ({\bf 0})\rangle & =& 
\frac{1}{\sqrt{6}}
\Bigg (\frac {R_{V}^2}{\pi} \Bigg)^{\frac{3}{4}}
\int d{\bf k} 
\exp\Big(-\frac {R_{V}^2 {\bf k}^2}{2}\Big)
{{Q_1}_r}^{i}({\bf k})
^\dagger u_r^\dagger \sigma^m 
\tilde {{Q_2}_s}^{i} 
(-{\bf k})v_s
|vac\rangle,
\label{kstr_phi}
\end{eqnarray}
with the quark-antiquark constituents for the vector meson
$V(\equiv {K^*}^+, {K^*}^0, \phi)$, as
given by $(Q_1,{\bar {Q_2}}) \equiv (u,\bar s), (d,\bar s), (s,\bar s)$.
The parameters $R_{P}$ and $R_V$ 
in equations (\ref{pseudoscalar_meson}) and (\ref{kstr_phi}),
correspond to the the harmonic oscillator strengths for the pseudoscalar 
and vector mesons. In the pseudoscalar state, 
$|P({\bf p})\rangle$ given by equation (\ref{pseudoscalar_meson}), 
$\lambda_1$ and $\lambda_2$ are the 
fractions of the mass (energy) of the pseudoscalar meson
at rest (in motion), carried by the constituent 
antiquark, ${\tilde {Q_2}}$ and the constituent quark,
$Q_1$. The constituent quark (and antiquark) of the hadron 
occupying specific energy level is similar to as in the MIT bag model 
\cite{MIT_bag}. The fractions of the energy carried by the 
constituent quark (antiquark) are calculated by assuming 
the binding energy of the meson as shared by the quark 
(antquark) to be inversely proportional to the quark 
(antiquark) mass \cite{spm782}. For the pion states, $\pi^+$ and $\pi^0$, 
which are light ($q=u,d$) quark-antiquark bound states, 
the fractions of energy carried by the quark and antiquark 
are the same, i.e., $\lambda_1=\lambda_2=1/2$, as the masses of the $u$
and $d$ quarks are assumed to be the same. 
The decay widths for the processes $K^* \rightarrow K\pi$
and $\phi \rightarrow K \bar K$, 
are obtained from the matrix element of the light quark
antiquark pair creation term of the free Dirac Hamiltonian density,
between the initial and the final states \cite{amspmwg}. 
The matrix element is multiplied with a strength parameter, 
$\gamma_{V(=K^*,\phi)}$, which is fitted to the observed
decay width of the vector meson in vacuum.

\subsubsection{\bf{DECAY WIDTH OF $K^*\rightarrow K \pi$}}
\noindent In the absence of the pseudoscalar-vector mesons
mixing, the decay width for $K^* \rightarrow K\pi$ 
is obtained from the matrix element of the free Dirac
Hamiltonian between the initial and final states as 
\cite{amspmwg}
\begin{eqnarray}
\Gamma\left(K^{*} ({\bf 0})\rightarrow 
K ({\bf p})+\pi ({-\bf p})\right)
= \gamma_{K^*}^2 g^2\frac{8\pi^2p_K^0p_\pi^0}{3m_{K^*}}A_{K^*}
(|\bfs p|)^2|\bfs p|^3,
\label{gammakstr}
\end{eqnarray}
where $p_K^0=(|{\bf p}|^2+m_K^2)^{1/2}$ and 
$p_\pi^0=(|{\bf p}|^2+m_\pi^2)^{1/2}$ are the energies of
the outgoing $K$ meson and pion respectively, in terms 
of the magnitude of the 3-momentum of $K(\pi)$ meson,
$|{\bf p}|$, given as
\begin{equation}
|\bfs p|=\Bigg (\frac{m_{K^{*}}^2}{4}-\frac{m_{K}^2+m_{\pi}^2}{2}
+\frac{\left(m_{K}^2-m_{\pi}^2\right)^2}{4m_{K^{*}}^2}
\Bigg )^{1/2}.
\label{mokp}
\end{equation}
In equation (\ref{gammakstr}), $A_{K^*}(|{\bf p}|)$ is given as
\begin{eqnarray}
A_{K^*}(|{\bf p}|)=6c_{K^*}
\Big(\frac{\pi}{a_{K^*}}\Big)^{{3}/{2}}
\exp\left[\Big (a_{K^*}b_{K^*}^2
-\frac{1}{2}\left(\lambda_2^2 R_K^2
+\frac{1}{4}R_\pi^2\right)\Big){|\bf p|}^2\right] 
\times 
\Big[{F_0}_{K^*}+\Big (\frac {3{F_1}_{K^*}}{2a_{K^*}}\Big )\Big],
\label{apkstr}
\end{eqnarray}
where,
\begin{eqnarray}
{F_0}_{K^*}&=&(b_{K^*}-1)\left(1-\frac{1}{8M_q^2}|{\bfs p}|^2
(\lambda_2-\frac{1}{2})^2\right) \nonumber \\
&-&(b_{K^*}-\lambda_2)\left(\frac{1}{2}+\frac{1}{4M_q^2}
{|\bfs p|}^2
\left(\frac{3}{4}b_{K^*}^2-\frac{5}{4}b_{K^*}
+\frac{7}{16}\right)\right)
\nonumber\\
&-&(b_{K^*}-\frac{1}{2})\left[\frac{1}{2}+\frac{1}{4M_q^2}{|\bfs p|}^2
\left(\frac{3}{4}b_{K^*}^2-(1+\frac{1}{2}\lambda_2)b_{K^*}
+\lambda_2-\frac{1}{4}\lambda_2^2\right)\right]
\label{c0kstr}
\\
{F_1}_{K^*}&=&-\frac{1}{4M_q^2}\left[\frac{5}{2}b_{K^*}-\frac{9}{8}
-\frac{11}{12}\lambda_2\right].
\label{c1kstr}
\end{eqnarray}
The parameters $a_{K^*}$, $b_{K^*}$, $c_{K^*}$ are given as
\begin{eqnarray}
a_{K^*}=\frac{\left(R_{K^*}^2+R_K^2+R_\pi^2\right)}{2},\;\;
b_{K^*}=\frac{1}{2a_{K^*}}\left(R_K^2\lambda_2
+\frac{1}{2}R_\pi^2\right),\;\;
c_{K^*}=\frac{1}{12\sqrt 3}
\left(\frac{R_{K^*}^2 R_K^2 R_\pi^2}{\pi^3}\right)^{{3}/{4}}.
\label{abckstr}
\end{eqnarray}
In equations (\ref{c0kstr})--(\ref{c1kstr}), $M_q$ is the constituent
light quark ($q=u,d$) mass.
As has already been mentioned, the expression for the decay
width for $K^* \rightarrow K\pi$ as given by equation
(\ref{gammakstr}) (modulo $\gamma_{K^*}^2$) is obtained  
from the matrix element of the free Dirac Hamiltonian
between the initial and final states. For $\pi_0$
($\pi^\pm$) in the final state, the value of 
$g^2$=1(2). The decay amplitude is multiplied by $\gamma_{K^*}$, 
which is the production strength of $K\pi$ from decay of $K^*$ meson, 
and is fitted from its vacuum decay width. 
The decay width has the dependence on the
masses of the decaying and outgoing mesons,
through $|{\bf p}|$, and its dependence
as given by equation (\ref{gammakstr}), is mainly through   
a polynomial part multiplied by an exponential part. 

The expression for the decay width of $K^*\rightarrow K\pi$ 
given by equation (\ref{gammakstr}) does not account for
the mixing of the $K$ and $K^*$ mesons in the
presence of the magnetic field. The mixing of the pseudoscalar
mesons and the vector mesons leads to a drop (increase)
in the mass of the $K$ meson (longitudinal component 
of the $K^*$ meson), given by equaion (\ref{mpv_long}). 
This leads to  
the expression for the decay width of $K^*\rightarrow K\pi$ 
to be modified to \cite{dmeson_PV_amspm}
\begin{eqnarray}
&&\Gamma^{PV}(K^* ({\bf 0}) \rightarrow  
K({\bf p}) {\pi} (-{\bf p}))
=\gamma_{K^*}^2 g^2\frac{8\pi^2}{3}
\Bigg [ 
\Bigg(\frac{2}{3} |{\bf p}|^3 
\frac {p^0_K (|{\bf p}|) p^0_{\pi}(|{\bf p}|)}{m_{K^*}}
A^{K^*}(|{\bf p}|)^2 \Bigg)
\nonumber \\
&+&\Bigg(\frac{1}{3} |{\bf p}|^3 
\frac {p^0_K(|{\bf p}|) p^0_{\pi}(|{\bf p}|)}{m_{K^*}^{PV}}
A^{K^*}(|{\bf p}|)^2 \Bigg) \Big({|{\bf p}|\rightarrow |{\bf p|}
(m_{K^*} = m_{K^*}^{PV}, m_{K} = m_{K}^{PV}
)}\Big)
\Bigg]. 
\label{gammakstr_mix}
\end{eqnarray}
The first term corresponding to the transverse
polarizations for the vector $K^*$ meson are
unaffected by the mixing of the $K$ and $K^*$
states, whereas the second term has the masses 
the $K$ and the longitudinal $K^*$ mesons modified 
due to the $K-K^*$ mixing.

\subsubsection{\bf {DECAY WIDTH OF $\phi\rightarrow K \bar K$}}
The decay width of $\phi \rightarrow K \bar K$ is calculated 
to be 
\begin{eqnarray}
\Gamma(\phi ({\bf 0})\rightarrow K ({\bf P})\bar K(-{\bf P}))
= \gamma_\phi^2\frac{8\pi^2}{3}|{\bf P}|^3
\frac {{P^0}_{K} {P^0}_{\bar K}}{m_{\phi}}
A_{\phi}(|{\bf P}|)^2
\label{gammaphikkbar}
\end{eqnarray}
in the absence of the pseudoscalar-vector mesons mixing
effects. 
In equation (\ref{gammaphikkbar}) for the expression
of decay width of $\phi$ to $K\bar K$, 
$P^0_{K(\bar K)}=\big(m_{K (\bar K)}^2
+|{\bf P}|^2\big)^{\frac{1}{2}}$, 
and, $|\bfs P|$ is the magnitude of the momentum of the outgoing 
$K(\bar K)$ meson, given as
\begin{equation}
|{\bf P}|=\Bigg (\frac{{m_\phi}^2}{4}-\frac {{m_K}^2+{m_{\bar K}}^2}{2}
+\frac {({m_K}^2-{m_{\bar K}}^2)^2}{4 {m_\phi}^2}\Bigg)^{1/2}.
\label{pkkbar}
\end{equation}
In the above, $A_{\phi}(|\bfs P|)$ is given as
\begin{equation}
A_{\phi}(|{\bf P}|)=6c_\phi\exp[(a_\phi {b_\phi}^2
-R_K^2\lambda_2^2)|{\bf P}|^2]
\cdot\Big(\frac{\pi}{a_\phi}\Big)^\frac{3}{2}
\Big[F_{0 \phi}+F_{1 \phi}\frac{3}{2a_\phi} \Big],
\label{ap}
\end{equation}
with
\begin{eqnarray}
&&F^{\phi}_0=(\lambda_2-1)-\frac{1}{2M_q^2}
|\bfs P|^2(b_{\phi}-\lambda_2)
\left(\frac{3}{4}b_{\phi}^2-(1+\frac{1}{2}\lambda_2)b_{\phi}
+\lambda_2-\frac{1}{4}\lambda_2^2\right),
\nonumber\\
&&F^{\phi}_1=\frac{1}{4 M_q^2}
\left[-\frac{5}{2}b_{\phi}+\frac{2}{3}
+\frac{11}{6}\lambda_2\right].
\label{c012phi}
\end{eqnarray}
and the parameters $a_\phi$, $b_\phi$ and $c_\phi$ given as
\begin{equation}
a_\phi=\frac{1}{2}R_{\phi}^2+R_K^2, \;\;
b_\phi=R_K^2\lambda_2/a_\phi,\;\;
c_{\phi}=\frac{1}{6\sqrt{6}}\cdot
\left(\frac{R_\phi^2}{\pi}\right)^{{3}/{4}}
\cdot\left(\frac{R_K^2}{\pi}\right)
^{{3}/{2}},
\label{abcphi}
\end{equation}

When the pseudoscalar meson--vector meson mixing is included,
the masses of the $K$ and $\bar K$ mesons and the longitudinal
component of the $\phi$ meson are modified due to $K-K^*$,
$\bar K-\bar {K^*}$ and $\phi-\eta'$ mixings.
The modified expression for the decay width
$\phi\rightarrow K\bar K$ is given as
\cite{charmonium_PV_amspm}
\begin{eqnarray}
&& \Gamma ^{PV}(\phi ({\bf 0})\rightarrow K ({\bf P})\bar K(-{\bf P}))
= \gamma_\phi^2\frac{8\pi^2}{3}
\Bigg [ \Bigg ( \frac{2}{3} |{\bf P}|^3
\frac {{P^0}_{K} {P^0}_{\bar K}}{m_{\phi}}
A_{\phi}(|{\bf P}|)^2 \Bigg)\nonumber \\
&+ & \Bigg ( \frac{1}{3} |{\bf P}|^3 
\frac {{P^0}_{K} {P^0}_{\bar K}}{m_{\phi}}
A_{\phi}(|{\bf P}|)^2 \Bigg)
\Big(|{\bf P}|\rightarrow 
|{\bf P}| (m_{\phi}=m_{\phi}^{PV},
m_{K(\bar K)}=m_{K(\bar K)}^{PV})\Big)
\Bigg]
\label{gammaphikkbar_mix}
\end{eqnarray}

\section{Results and Discussions}

\begin{figure}
\includegraphics[width=16.4cm,height=16.4cm]{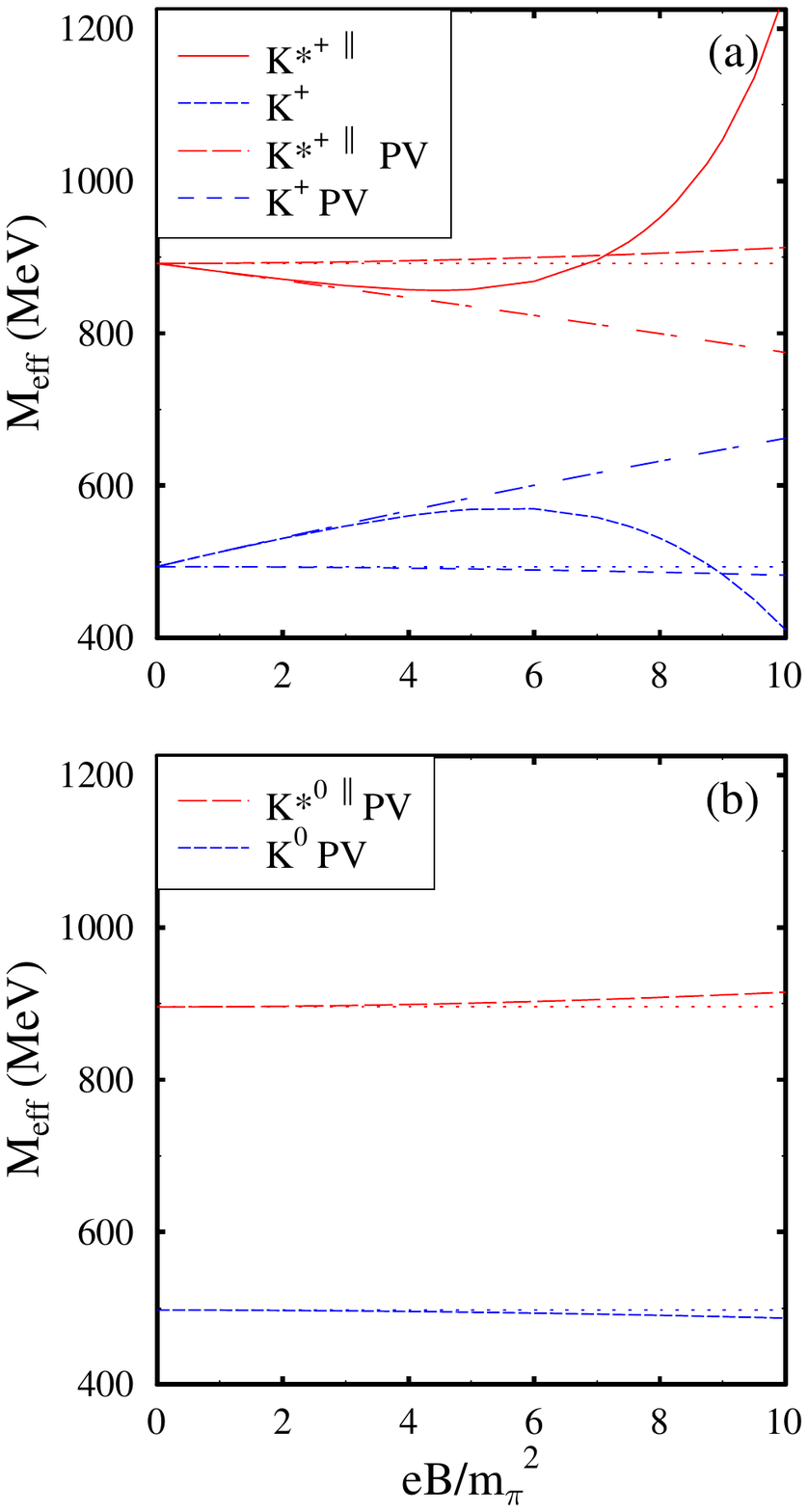}
\caption{(Color online)
The masses of the $K$ and the longitudinal components 
of the $K^*$ mesons are plotted as functions of $eB/{m_\pi^2}$. 
The effects of the pseudoscalar--vector (PV) mesons mixing on these
masses are shown for the charged and neutral mesons in panels
(a) and (b) respectively. The Landau contributions to the masses
of the charged $K^+$ and ${K^*}^+$ mesons are shown as the 
dot-dashed lines.
}
\label{mKKstr}
\end{figure}

\begin{figure}
\includegraphics[width=16.4cm,height=16.4cm]{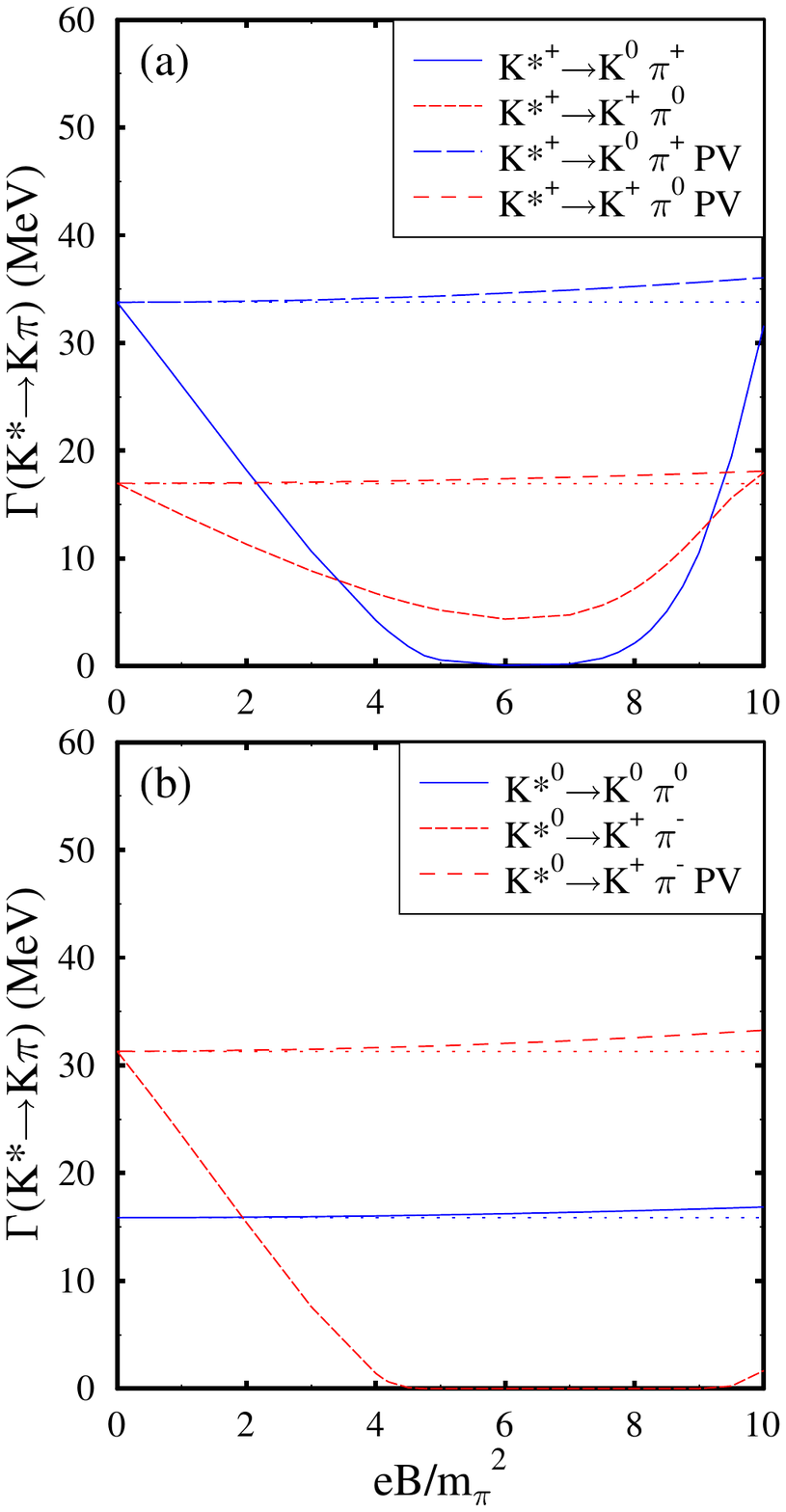}
\caption{(Color online)
The decay widths for $K^*\rightarrow K\pi$
for the charged ${K^*}^+$ and neutral ${K^*}^0$ are 
plotted as functions of $eB/{m_\pi^2}$
in panels (a) and (b) respectively. These are shown 
including both the pseudoscalar--vector mesons (PV) mixing 
as well as Landau level contribtuions 
for the charged mesons.
These are also shown when the mixing effects (indicated
as PV) are only taken into account. 
}
\label{dwFT_Kstr}
\end{figure}

\begin{figure}
\includegraphics[width=16.4cm,height=16.4cm]{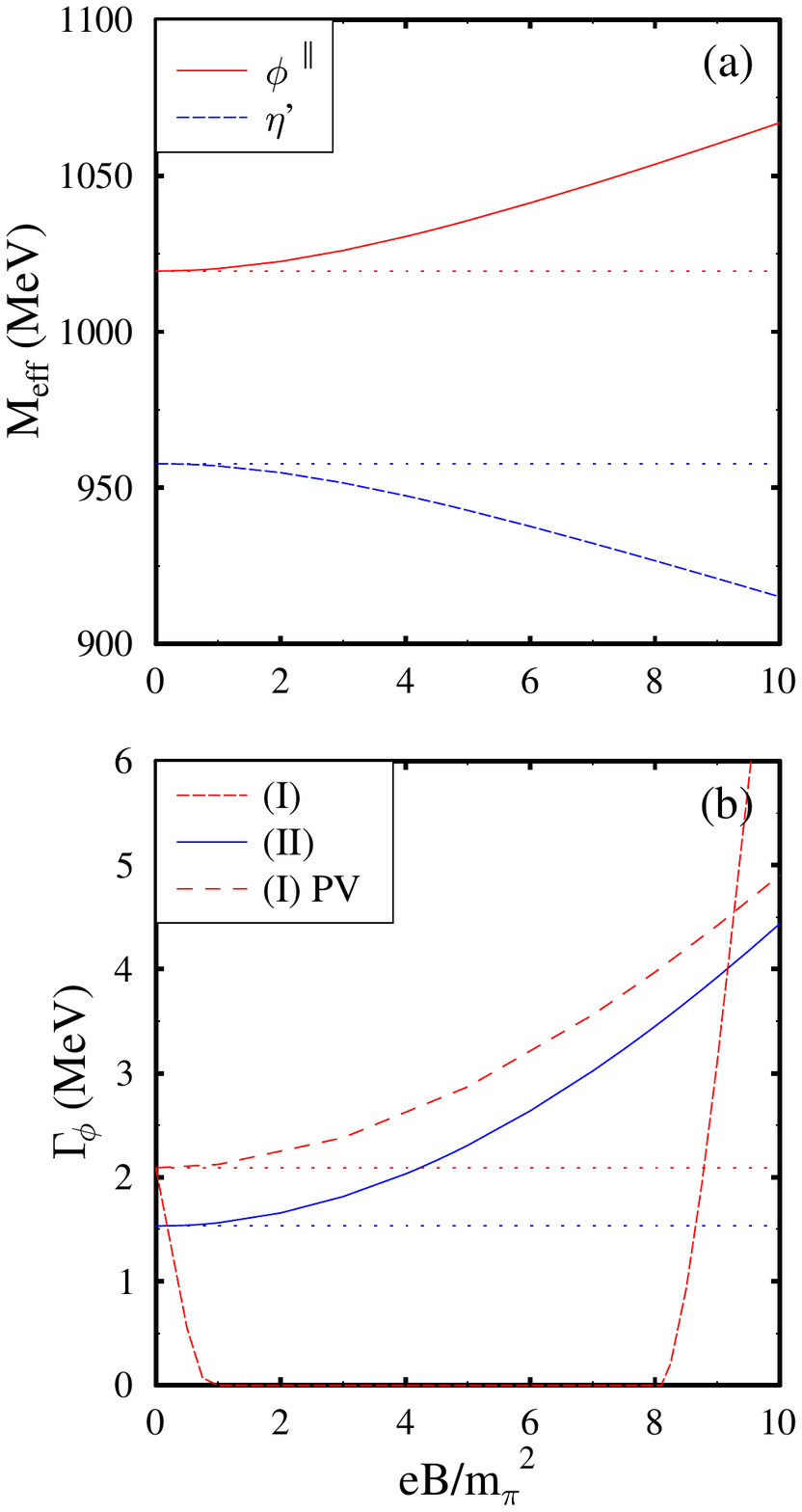}
\caption{(Color online)
The mass modifications of the $\phi$ and $\eta '$ mesons
arising from the pseudoscalar--vector meson mixing are shown 
as functions of  $eB/{m_\pi^2}$ in panel (a).
Panel (b) shows the effects of the magnetic field
on the partial decay widths of $\phi$
to (I) $K^+ K^-$ and (II) $K^0 {\bar K}^0$ respectively.
The Landau contributions to the masses of the charged
mesons $K ^ \pm $ are taken into account and these are
compared to the case of when these effects are not considered.
}
\label{mass_dwFT_phi}
\end{figure}

In the presence of a uniform magnetic field, 
the modifications of the masses of the $K$, $K^*$, 
$\phi$ and $\eta '$ due to the mixing of the 
pseudoscalar and vector mesons are investigated.
These are in addition to the  
Landau level contributions for the charged mesons. 
The mixing is taken into account through 
a phenomenological Lagrangian
interaction given by equation (\ref{PVgamma}). 
The coupling strength parameter $g_{PV}$ 
for the radiative decay of the 
vector meson, $V$ to the pseudoscalar meson, $P$,
described by the interaction Lagrangian
(\ref{PVgamma}) is determined from the observed
decay width of $V\rightarrow P\gamma$ in vacuum.
For the processes  ${K^*}^+ \rightarrow K^+ \gamma$, 
${K^*}^0 \rightarrow K^0 \gamma$ and 
${\phi} \rightarrow \eta ' \gamma$, the observed decay
widths of 50.292 keV, 46.827 keV and 0.26429 keV
\cite{pdg_2019_update} determine the coupling 
parameters, $g_{{K^+}{K^*}^+}$,  $g_{{K^0}{K^*}^0}$, 
and $g_{{\eta '}{\phi}}$ to be 0.5793, 0.5611 and 0.7043
respectively. The vacuum values for the masses (in MeV) of these mesons
are taken to be  $m_{{K^*}^+}=891.66,\,\,m_{{K^*}^0}=895.55,\,\,
m_{{K}^+}=493.677,\,\,m_{{K}^0}=497.61,\,\,m_{\phi}=1019.461,\,\,
m_{\eta '}=957.78$ \cite{pdg_2019_update}.
The presence of a magnetic field
leads to the mixing of the pseudoscalar meson
and the longitudinal component of the vector meson,
with their modified masses given by equation 
(\ref{mpv_long}), due to the mixing effect.
As has already been mentioned, for the charged mesons, 
these modifications in the masses are in addition 
to the contribution arising from the lowest Landau levels, 
due to the direct interactions of the charged mesons with the
external magnetic field. The masses of the 
charged and neutral open strange mesons
are plotted in figure \ref{mKKstr}. 
The mixing of the pseudoscalar and vector mesons 
(indicated as PV) are observed to be a monotonic drop (rise) in the
mass of the $K^+(K^0)$ (longitudinal component of the vector
${K^*}^+({K^*}^0$)) meson. 
The modifications of the masses of the charged as well as neutral 
$K$ and $K^*$ mesons are observed to be quite small 
due to the (PV) mixing effects, with modified values of 
for ${K^*}^+$,  ${K}^+$,  ${K^*}^0$,  ${K}^0$ masses as 
897 (912.46), 490.75 (482.4), 900.49 (914.87)
and 494.88 (487.1) mesons respectively, 
at $eB=5 (10) m_\pi^2$.
For the maximum value of the magnetic field considered
in the present work, $eB=10 m_\pi^2$, the mass shifts (in MeV) for 
${K^*}^+$,  ${K}^+$,  ${K^*}^0$ and ${K}^0$ mesons as 
thus observed to be around 21, 11, 19 and 10.5 respectively.  
The Landau level contributions are observed to lead to 
a monotonic drop (rise) in the mass of the ${K^*}^+$ ($K^+$) 
with increase in magnetic field.
From panel (a) of figure \ref{mKKstr}, it is observed that 
there is a drop (increase) in the mass of the ${K^*}^+$(${K}^+$) meson 
with increase in the magnetic field, upto a value of $eB$
of the order of around $5m_\pi^2$, when the Landau contributions dominate 
over the mixing effects. 
As the mass difference of these mesons becomes smaller
for larger values of the magnetic field,
the mixing effect is observed to become more appreciable
and this starts dominating over the contributions from the 
Landau levels. This is observed to lead to quite a dominant 
rise (drop) of the mass of the ${K^*}^+$ ($K^+$) meson 
for $eB$ greater than about $6 m_\pi^2$.
The mass modifications of the charged $K$ and $K^*$
mesons are thus observed to be much more pronounced
as compared to the neutral mesons, due to the 
additional effects from the Landau level contributions.
These mass modifications of the charged open strange mesons
are observed to modify appreciably the decay widths 
of ${K^*}^+ \rightarrow K\pi$, as can be seen from
figure \ref{dwFT_Kstr}. 
The mass modifications of the neutral mesons, ${K^*}^0$ and
$K^0$ due to the magnetic field, which arise only due to
the pseudoscalar-vector mesons mxing, are shown in panel (b)
of figure \ref{mKKstr}. These mass changes are observed 
to be moderate due to the small mixing coupling parameter, 
$g_{K^0{K^*}^0}$=0.5611.
It might be worthwhile to compare the effects of the
mixing on the masses of the open strange mesons as studied in the
present work, with the mass modifications 
of the open charm mesons due to the mixing effects
\cite{dmeson_PV_amspm}. The total width of the neutral
$D^*$ meson is not yet measured experimentally, but the
branching raito of the two modes ${D^*}^0\rightarrow D^0 \pi^0$
and ${D^*}^0\rightarrow D^0 \gamma$ is measured to be
64.7:35.3 \cite{pdg_2019_update}. 
The decay width of ${D^*}^0\rightarrow D^0 \pi^0$
(and hence of ${D^*}^0\rightarrow D^0 \gamma$) 
is obtained \cite{Gubler_D_mag_QSR,dmeson_PV_amspm} 
by taking the coupling strengths
of the decays ${D^*}^0\rightarrow D^0 \pi^0$ and
${D^*}^+\rightarrow D^+ \pi^0$ to be the 
same. This is observed to 
lead to the pseudoscalar--vector mesons mixing parameter
for the neutral open charm mesons, $g_{D^0{D^*}^0}$ to be
quite large, about 4 times larger than the mixing
parameter for ${D^*}^+ - D^+$ mixing
\cite{Gubler_D_mag_QSR,dmeson_PV_amspm}. The mass
modifications for the neutral open charm mesons
due to the mixing effects
are thus observed to be much more pronounced compared
to the mass modifications of the charged $D$ and $D^*$ mesons.
On the other hand, in the present investigation,
the mass modifications of the neutral open strange mesons 
are moderate. The much larger value of the mixing parameter
in the neutral open charm sector could be due to the availability
of a single channel for $D^*\rightarrow D\pi$, 
which is ${D^*}^0 \rightarrow D^0 \pi$, in addition to the
radiative decay channel ${D^*}^0 \rightarrow D^0 \gamma$,
and these are the only decay modes of ${D^*}^0$ meson.
The radiative decay width is comparable to the 
decay width of $D^*\rightarrow D\pi$ for the neutral $D^*$.
On the other hand, for the open strange meson sector, there are two channels
${K^*}^0 \rightarrow K^0 \pi^0$ and ${K^*}^0 \rightarrow K^+ \pi^-$, 
and the decay width of ${K^*}^0 \rightarrow (K\pi)^0$ is about 
99.754 \% of the total width of ${K^*}^0$ meson. 
The radiative decay wdith of ${K^*}^0 \rightarrow K^0 \gamma$
is about 0.246 \% of its total width, which makes the mixing
of ${K^*}^0-K^0$ to be quite small. Also, for the charged
${K^*}^+$ meson, the decay is dominated by ${K^*}^+ \rightarrow (K\pi)^+$
and the decay to $K^+\gamma$ is around $0.099 \%$ of its
total width. These lead to the mass modifications of the
$K$ and $K^*$ (both the charged and neutral mesons)
due to $K$-$K^*$ mixings to be quite moderate, as can be seen 
in figure \ref{mKKstr}. The masses of the charged open strange
mesons, ${K^*}^\pm$ and $K^\pm$ mesons, as modified due to 
the Landau level contributions are $({m_{{K^*}^\pm}^2-eB})^{1/2}$ 
and $({m_{K^\pm}^2+eB})^{1/2}$ respectively \cite{Chernodub}. 
The mass of the charged vector meson, $V$, becoming negative above
a critical magnetic field, $(eB)_{crit}=m_V^2$ leads to
condensation of the charged vector meson \cite{Chernodub}.  
For magnetic fields higher than the critical magnetic field,
$(eB)_{crit}=m_\rho^2 \sim 30 m_\pi^2$, there is condensation
of the charged $\rho$ mesons ($u\bar d (d\bar u)$ bound states),
arising from the gluon mediated attractive interaction 
of the quark and antiquark of different flavours 
in spin 1 state \cite{Chernodub}. For still stronger
magnetic fields, $eB$ greater than
$(eB)_{crit}=m_{K^*}^2 \simeq 40 m_\pi^2$, 
there should be condensation of the charged $K^*$ mesons 
($u \bar s (s \bar u)$ bound states) as well. 

The decay widths, $K^* \rightarrow K \pi$ and $\phi \rightarrow 
K\bar K$ as modified due to the mass modifications 
of these mesons in the presence of a magnetic field are studied 
using a field theoretical model of composite hadrons
with constituent quarks and antiquarks.
These decay processes are studied using a light quark-antiquark
pair creation term, which is the quark-antiquark creation
term of the free Dirac Hamiltonian in terms of the
constituent quark fields. The matrix element of this term
between the initial and final states of the decay process
is calculated to compute the decay widths of the 
processes $K^* \rightarrow K \pi$ and $\phi \rightarrow 
K\bar K$. As has already been mentioned, 
$\lambda_1$ and $\lambda_2$ in the state $K(K^+,K^0)$ meson 
($q \bar s$ bound state, $q=(u,d)$) 
given by equation (\ref{pseudoscalar_meson}), 
are the fractions of the mass (energy) of the $K$ meson
at rest (in motion) carried by the constituent strange antiquark
and the constituent light quark ($u$, $d$), 
and $\lambda_1+\lambda_2$=1. 
These are calculated by assuming the binding energy of the hadron 
as shared by the quark (antquark) to be inversely 
proportional to the quark (antiquark) mass
\cite{spm782}. With the vacuum mass (in MeV) of $K^+ (K^0)$ meson 
to be given as 493.677 (497.61), $M_{u,d}$=330 MeV \cite{amspmwg}, 
and $M_s$=480 MeV, the value of $\lambda_2$ is obtained as 
0.71.

The harmonic oscillator strengths of the pseudoscalar mesons
($K$, $\pi$) and vector mesons ($\phi$ and $K^*$) are needed to
calculate the decay widths of $K^* \rightarrow K\pi$ and
$\phi \rightarrow K \bar K$. The value of $R_\pi$=(211 MeV)$^{-1}$ 
\cite{spm782,amspmwg} 
was fitted from the value of the charge radius squared of pion given as
$0.4 {\rm fm}^2$. We determine the harmonic oscillator strength
parameter for the $K$ meson, $R_K$, assuming the ratio $R_K/R_\pi$
to be same as the ratio of their charge radii, $({r_{ch}})_K/({r_{ch}})_\pi$.
Taking $({r_{ch}})_K$=0.56 fm \cite{pdg_2019_update}, the 
value of $R_K$ is obtained as (238.3 MeV)$^{-1}$.
The value of $R_\phi$ is obtained from the observed decay width
of $\phi \rightarrow e^+e^-$ of 1.26377 keV. The expression
of the decay width is given as \cite{Van_Royen}
\begin{equation}
\Gamma (\phi \rightarrow e^+ e^-)
=\frac {16 \pi \alpha ^2}{9 m_\phi^2} |\psi({\bf r}={\bf 0})|^2,
\end{equation}
with $\alpha=1/137$, yields the value if $R_\phi$ 
to be (290.7 MeV)$^{-1}$. 
The value of $R_{K^*}$ is assumed to be same as $R_K$
in the present work. 
The values of $\gamma_{K^*}$ for the decays 
${K^*}^+\rightarrow (K\pi)^+$ and
${K^*}^0\rightarrow (K\pi)^0$,
as fitted to their observed vacuum decay widths 
of 50.75 and 47.18 MeV \cite{pdg_2019_update}, 
are obtained as 2.4742 and 2.347 respectively. 
These yield the values of the
decay widths for the subchannels of the 
${K^*}^+$ meson to $K^+\pi^0$ and $ K^0 \pi^+$ to be
16.98 and 33.77 MeV and of decay widths of ${K^*}^0$ to 
$K^0\pi^0$ and $K^+ \pi^-$ to be 15.87 and 31.31 MeV
respectively. The effects of the magnetic field
on the decay widths of the charged
and neutral $K^*\rightarrow K\pi$ are plotted
in figure \ref{dwFT_Kstr}.
The decay widths for ${K^*}^+$
to  $K^0 \pi^+$ as well as to  $K^+ \pi^0$ are observed
to increase monotonically with the magnetic field, 
when the mass modifications are taken into account
only due to mixing of the $K$ and $K^*$ mesons
(indicated as PV in figure \ref{dwFT_Kstr}), 
and these modifications are seen to be moderate.
The vacuum values of these decay wdiths of 33.77
and 16.98 are observed to be modified to 40.63
and 20.34 at $eB=10 m_\pi^2$ for ${K^*}^+\rightarrow K^0\pi^+$
and ${K^*}^+\rightarrow K^+\pi^0$ respectively.
When the Landau level contributions are
also taken into account for the charged mesons,
${K^*}^+$ and $\pi^+$, there is observed to be
a drastic drop in the decay width of  
${K^*}^+ \rightarrow K^0 \pi^+$, mainly due to
the drop (rise) in mass the charged vector (pseudoscalar)  
meson from Landau level contributions. The value of the decay width
vanishes at around $eB = 5 m_\pi^2$ and remains
zero upto around $eB = 7 m_\pi^2$. As the magnetic
fueld is further increased, the 
dominant increase in mass of ${K^*}^+$ due to mixing effects,
leads to an increase in the decay width of 
${K^*}^+ \rightarrow K^0 \pi^+$. The decay width of 
${K^*}^+ \rightarrow K^+ \pi^0$ is observed to have a
similar behaviour when the Landau level contributions
as well as mixing effects are taken into account.

The masses of the $\phi$ and $\eta'$ due to their mixing
on the presence of a magnetic field are shown in panel (a)
in figure \ref{mass_dwFT_phi}. There is observed to
be an increase (drop) in $\phi (\eta')$ meson mass 
due to this mixing. There are no Landau level contributions
to their masses as these mesons are charge neutral.
The values of $\gamma_\phi$ for the decay of 
$\phi \rightarrow K^+K^-$ and $\phi \rightarrow K^0 {\bar {K^0}}$
are obtained as 2.38 and 2.41 are obtained,
as fitted from their observed vacuum decay widths
of 2.0905 and 1.53 MeV \cite{pdg_2019_update}.
The decay width of  $\phi$ to $K^+ K^-$ as well as to
the neutral $K^0 \bar {K^0}$ are shown in panel (b)
in figure \ref{mass_dwFT_phi}. The effects of the 
magnetic field on the decay width
to the neutral $K\bar K$ pair arise from the mass modifications
of the $\phi$ meson as well as of the $K^0$ (and $\bar {K^0}$)
mesons due to the $\phi-\eta'$, ${K^*}^0-{K^0}$ 
(and $\bar {K^*}^0-\bar {K^0}$) mixings. These mixing effects
lead to a rise of decaying $\phi$ mass and drop in the
outgoing pseudoscalar $K^0$ and ${\bar {K^0}}$ meson masses.  
These mass modifications are observed as a monotonic 
increase in the decay width of $\phi\rightarrow K^0 {\bar {K^0}}$
with increase in the magnetic field. 
The decay width of $\phi\rightarrow K^+ K^-$
also shows a similar trend when the Landau
contributions to the masses of these charged
kaons and antikaaons are not taken into consideration.
The Landau level contributions lead to increase
in the masses of the charged $K\bar K$,
due to which it is observed that the decay
width vanishes at around $eB=m_\pi^2$
and remains zero upto around  $eB=8.1 m_\pi^2$.
As the magnetic field is increased
further, the decay to $K^+K^-$ is observed to 
again become kinematically possible, as the masses 
of the charged $K$ and $ \bar K$ decrease at high
magnetic field due to the dominance of the mixing 
($K^\pm - {K^*}^\pm$) effects. This 
can be seen in figure \ref{mKKstr}
for the ${K^*}^+ - K^+$ mixing in the presence
of a magnetic field. 

The present work of study of effects of strong magnetic fields
on the strange vector mesons is complementary to the recent work
on the study of the spectral functions and the production 
cross-sections of the vector strange mesons ($\phi$, $K^*(\bar {K^*})$)
from $K\bar K$, $K(\bar K)\pi$ scatterings, 
in isospin asymmetric strange hadronic matter 
\cite{strangedecaywidths}. 
It may be mentioned here that 
in the relativistic heavy ion collision experiments, 
contrary to the heavy flavour mesons which 
are produced at the early stage when the magnetic field 
produced can still be high, the strange vector mesons, 
$\phi$ and $K^*$ are produced at the later stages of evolution
when the magnetic field is expected to be very small. 
Hence it is not feasible to study the effects of strong 
magnetic fields on these strange mesons from the experimental 
observables of heavy ion collision experiments. 
However, apart for being a topic of of theoretical interest,
the study of the effects of magnetic field on the properties 
of the strange mesons, e.g, the kaons and $\phi$ mesons, 
could have implications on the composition of 
neutron star matter due to possible existence
of large magnetic fields in the interior 
of magnetars \cite{Aguirre_neutral_K_phi}. 

\section{Summary}

In the present work, we have studied the modifications of the
masses of the vector mesons ($\phi$ and $K^*$) and the 
pseudoscalar mesons ($\eta'$, $K$) in the presence of 
strong magnetic fields, due to $\phi-\eta'$, $K^*-K$ mixings. 
The modifications in masses of the charged mesons in their ground sttes 
are in addition to the lowest Landau level contributions.
The effects of the mass modifications due to the magnetic field
on the decay widths of $K^*\rightarrow K\pi$ and $\phi\rightarrow
K \bar K$ are investigated using a field theoretic model of composite
hadrons. For the neutral open strange mesons, $K^0$ and ${K^*}^0$,
as well as charged mesons $K^+$ and ${K^*}^+$, there is 
observed to be marginal modifications of the masses  leading to 
a drop (increase) in the mass of the pseudoscalar 
(longitudinal componenent of the vector) meson. 
The contributions of the Landau levels for the charged mesons
are observed to lead to a rise (drop) in the mass of 
$K^+ ({K^*}^+)$ meson, which gives rise to a smaller
mass difference of the masses of $K^+$ and ${K^*}^+$
with increase of the magnetic field. As the magnetic field 
is increased further, the mixing effects become more important, 
leading to a dominant increase (drop) in the ${K^*}^+$(${K}^+$) mass.
This leads to the effect of the magnetic field on the decay 
width of the charged $K^*$ meson to $K\pi$ 
(${K^*}^+\rightarrow K^0 \pi^+$ as well as ${K^*}^+\rightarrow K^+ \pi^0$) 
to have an initial drop followed by moderate change and then
a sharp rise when $eB$ is further increased.
The decay width ${K^*}^+\rightarrow K^0 \pi^+$ is observed to be zero 
for values of $eB/m_\pi^2$ between 5 and 7, due to
the increase in the mass of the charged pion in the final state.
The decay width of ${K^0}^* \rightarrow K^+ \pi^-$ is
observed to have a sharp drop with increase
in the magnetic field and becomes zero at around
$eB=4.2 m_\pi^2$. This is due to the positive contributions
from the Landau levels to the masses of both the charged mesons 
in the final state.
For the decay width of $\phi \rightarrow K \bar K$, 
the Landau contributions are observed to lead to
vanishing of the decay to the charged $K^+K^-$ pair 
above the value of $m_\pi^2$ for $eB$. The value of the decay width
for the charged $K\bar K$ remains zero upto around 8$m_\pi^2$.
On the other hand, the decay wdith of $\phi \rightarrow K^0 
{\bar {K^0}}$ is observed to increase monotonically
with the increase in the magnetic field. 

\begin{section}*{Acknowledgements}
One of the authors (AM) is grateful to ITP, University of Frankfurt,
for warm hospitality when this work was initiated. 
AM also acknowledeges financial support
from Department of Science and Technology (DST),
Government of India (project no. CRG/2018/002226).
\end{section}



\begin{thebibliography}{}

\bibitem{Tolos_Prog_Part_Nucl_Phys_112_103770_2020} 
L. Tolos, L. Fabbietti, Prog. Part. Nucl. Phys. {\bf 112},
103770 (2020).
\bibitem{C_Hartnack_Phys_Rep_510_2012_119} 
C. Hartnack, H. Oeschler, Y. Leifels, E. L. Bratkovskaya,
J. Aichelin, Phys. Rep. {\bf 510}, 119 (2012).

\bibitem{kaplan}
D.B. Kaplan and A. E. Nelson, Phys. Lett. B {\bf 175}, 57 (1986);
A. E. Nelson and D. B. Kaplan, ibid, {\bf 192}, 193 (1987). 

\bibitem{Glendenning_Schaffner_PRC_60_025803_1999}
N. K. Glendenning and J. Schaffner-Bielich, Phys. Rev.
C {\bf 60}, 025803 (1999).

\bibitem{Debades_PRC86_045803_2012}
S. Banik, R. Nandi, D. Bandyopadhyay, Phys. Rev. C {\bf 86},
045803 (2012).

\bibitem{Serot_QHD}
B. D. Serot, Rep. Prog. Phys. {\bf 55} 1855 (1992).


\bibitem{Krein_Prog_Part_Nucl_Phys}
G. Krein, A. W. Thomas, K. Tsushima, Prog. Part. Nucl. Phys. {\bf 100},
161 (2018).

\bibitem{qmc_phi}
K. Tsushima, K. Saito, A.W. Thomas, S.V. Wright, Phys. Lett. B
{\bf 429}, 239 (1998); ibid, Phys. Lett. E {\bf 436}, 453 (1998);
K. Tsushima, A. Sibirtsev, A.W. Thomas, Phys. Rev. C {\bf 62}
064904 (2000); ibid, J. Phys. G {\bf 27}, 34 (2001).


\bibitem{Oset_RamosNPA635_1998_99}
E. Oset and A. Ramos, Nucl. Phys. A {\bf 635}, 99 (1998).

\bibitem{Oset_Ramos_NPA_671_481_2000}
A. Ramos and E. Oset, Nucl. Phys. A {\bf 671}, 481 (2000).

\bibitem{Elena_16}

A. Ilner, D. Cabrera, C. Markert and E. Bratkovskaya,
Phys. Rev. C {\bf 95}, 014903 (2017).


\bibitem{Elena_17}
A. Ilner, J. Blair, D. Cabrera, C. Markert and E. Bratkovskaya,
Phys. Rev. C {\bf 99}, 024914 (2019).

\bibitem{STAR_Kstr} 
J. Adams et al [STAR Collaboration], Phys. Rev. C {\bf 71},
064902 (2005); M. M. Aggrawal et al [STAR Collaboration],
Phys. Rev. C {\bf 84}, 034999 (2011); L. Kumar [STAR Collaboration],
EPJ Web. Conf. {\bf 97}, 00017 (2015).

\bibitem{ALICE_Kstr} 
B. B. Abelev et al [ALICE Collaboration], Phys. Rev. C {\bf 91},
024609 (2014).

\bibitem{paper3}
 	P. Papazoglou, D. Zschiesche, S. Schramm, J. Schaffner-Bielich,
	H. St\"ocker, and W. Greiner, Phys. Rev. C {\bf 59},  411  (1999).

\bibitem{kmeson1}
     A. Mishra, E. L. Bratkovskaya, J. Schaffner-Bielich, S. Schramm
     and H. St\"ocker, Phys. Rev. C {\bf 70}, 044904 (2004).
\bibitem{isoamss}
A. Mishra and S. Schramm, Phys. Rev. C {\bf 74}, 064904 (2006).
\bibitem{isoamss1}
A. Mishra, S. Schramm and W. Greiner, Phys. Rev. C {\bf 78}, 024901 (2008).
\bibitem{isoamss2}
A. Mishra, A. Kumar, S. Sanyal, S. Schramm,
Eur. Phys. J A {\bf 41}, 205 (2009).

\bibitem{strangedecaywidths}
A. Mishra and S. P. Misra, arXiv:2006.10596 (nucl-th).

\bibitem{Tuchin_Review_Adv_HEP_2013}
K. Tuchin, Adv. High Energy Physics 2013, 490495 (2013).

\bibitem{magnetars}
K. Makishima. T. Mihara, F. Nagase and Y. Tanaka,
Astrophys. J. {\bf 525}, 978 (1999);
A. Melatos, Astrophys. Jour. {\bf 519}, L77 (1999). 

\bibitem{Kharzeev_NPA803_227_2008}
D. E. Kharzeev, L. D. McLerran and H. J. Warringa, 
Nucl. Phys. A {\bf 803}, 227 (2008).
\bibitem{Skokov_Illarionov_Tonnev_IJMPA24_5925_2009}
V. V. Skokov, A. Yu. Illarionov and V. D. Toneev,
Int. Jour. Mod. Phys. A {\bf 24}, 5925 (2009).
\bibitem{Kharzeev_PRD78_074033_2008} 
K. Fukushima, D. E. Kharzeev and H. J. Warringa,
Phys. Rev. D {\bf 78}, 074033 (2008). 
\bibitem{STAR_CME}
S. Voloshim for the STAR Collaboration, Nucl. Phys. A
{\bf 830}, 377c (2009); 
The STAR Collaboration, B. I. Abelev
et al, Phys. Rev. Lett. {\bf 103}, 251601 (2009);
STAR Collaboration, B. I. Abelev
et al, Phys. Rev. C {\bf 81}, 054908 (2010).

\bibitem{Voronyuk}
V. Voronyuk, V. D. Toneev, W. Cassing, E. L. Bratkovskaya,
V. P. Konchakovski and S. A. Voloshin,
Phys. Rev. C {\bf 83}, 54911 (2011).
\bibitem{Ajit_MHD}
A. Das, S. S. Dave, P. S. Saumia, A. M. Srivastava,
Phys. Rev. C {\bf 96}, 034902 (2017).

\bibitem{Aguirre_light_mesons}
R. M. Aguirre, Phys. Rev. D {\bf 96}, 096013 (2017).
\bibitem{Aguirre_neutral_K_phi}
R. M. Aguirre, Eur. Phys. Jour. A {\bf 55}, 2 (2019).
\bibitem{Pradip_rho_BT}
S. Ghosh, A. Mukherjee, M. Mandal, S. Darkar and P. Roy,
Phys. Rev. D {\bf 96}, 116020 (2017).

\bibitem{Chernodub}
M.N. Chernodub, Lect. Notes Phys. {\bf 871}, 143 (2013);
M. N. Chernodub, Phys. Rev. D {\bf 82}, 085011 (2010).

\bibitem{Hosaka_Prog_Part_Nucl_Phys} 
A. Hosaka, T. Hyodo, K. Sudoh, Y. Yamaguchi, S. Yasui,
Prog. Part. Nucl. Phys. {\bf 96}, 88 (2017).


\bibitem{charmonium_mag_QSR} S. Cho, K. Hattori, S. H. Lee, K. Morita
and S. Ozaki, Phys. Rev. Lett. {\bf 113}, 122301 (2014).
\bibitem{charmonium_mag_lee} 
S. Cho, K. Hattori, S. H. Lee, K. Morita
and S. Ozaki, Phys. Rev. D {\bf 91}, 045025 (2015).

\bibitem{Gubler_D_mag_QSR} P. Gubler, K. Hattori, S. H. Lee, M. Oka,
S. Ozaki and K. Suzuki, Phys. Rev. D {\bf 93}, 054026 (2016).

\bibitem{Alford_Strickland_2013}
J. Alford and M. Strickland, Phys. Rev. D {\bf 88}, 105017
(2013).

\bibitem{Suzuki_Lee_2017}
K. Suzuki and S. H. Lee, Phys. Rev. C {\bf 96}, 035203 (2017).

%

\bibitem{charmonium_PV_amspm}
Amruta Mishra and S. P. Misra,
 Phys. Rev. C {\bf 102}, 045204 (2020).
\bibitem{dmeson_PV_amspm}
Amruta Mishra and S. P. Misra,
arXiv:2005.00354 (hep-ph). 

\bibitem{spm781} S. P. Misra, Phys. Rev. D {\bf 18}, 1661 (1978).
\bibitem{spm782} S. P. Misra, Phys. Rev. D {\bf 18}, 1673 (1978).

\bibitem{amspmwg}
Amruta Mishra, S. P. Misra and W. Greiner, Int. J. Mod. Phys.
E {\bf 24}, 155053 (2015).

\bibitem{amspm_upsilon}
Amruta Mishra and S. P. Misra, Phy. Rev. C {\bf 95}, 065206 (2017).

\bibitem{friman}
B. Friman, S. H. Lee and T. Song, Phys. Lett, B {\bf 548}, 153
(2002).
\bibitem{amarvepja}
Arvind Kumar and Amruta Mishra, Eur. Phys. Jour.
A {\bf 47}, 164 (2011).

\bibitem{3p0}
E.S.Ackleh, T. Barnes and E. S. Swanson, Phys. Rev. D {\bf 54},
6811, 1996.
\bibitem{3p0_1}
T. Barnes, F. E. Close, P. R. Page and E. S. Swanson,
Phys. Rev. D {\bf 55}, 4157 (1997).

\bibitem{spmdiffscat} S. P. Misra and L. Maharana, Phys. Rev. D
{\bf 18}, 4103 (1978).


\bibitem{MIT_bag}
A. Chodos, R. L. Jaffe, K. Johnson and C. B. Thorn, 
Phys. Rev. D {\bf 10}, 2599 (1974).




\bibitem{pdg_2019_update}
M. Tanabashi et al (Particle Data Group), Phys. Rev. D
{\bf 98}, 030001 (2018) and 2019 update.


\bibitem{Van_Royen}
R. Van Royen and V. F. Weisskopf, Nuovo Cimento {\bf 28}, 131
(1967).


%
%

\end{thebibliography}
\end{document}